\documentclass[aps,prl,twocolumn,superscriptaddress,groupedaddressl]{revtex4} 
\usepackage{graphicx}  
\usepackage{float}
\pdfoutput=1
\usepackage{enumitem}
\usepackage{natbib}  
\usepackage{dcolumn}   
\usepackage{bm}        
\usepackage{amssymb}   
\usepackage{amsmath, amsfonts}   
\usepackage{mhchem} 
\usepackage{tabularx, multirow, graphicx}
\newcommand{\tilt}[2][2cm]{\rotatebox[origin=r]{90}{\parbox{#1}{\raggedleft #2}}}
\usepackage{xcolor,soul}
\usepackage[export]{adjustbox}
\usepackage{xcolor}
\usepackage[letterpaper,top=2cm,bottom=2cm,left=3cm,right=3cm,marginparwidth=1.75cm]{geometry}

\DeclareGraphicsExtensions{.eps}
\usepackage{epstopdf}

\begin{document}

\author{S. \surname{Shekarabi} }
\affiliation{%
    Grundlagen von Energiematerialien, Institut f\"ur Physik, Technische Universit\"at Ilmenau, 98693 Ilmenau, Germany
}%

\author{M. A. \surname{Zare Pour}}
\affiliation{%
    Grundlagen von Energiematerialien, Institut f\"ur Physik, Technische Universit\"at Ilmenau, 98693 Ilmenau, Germany
}%

\author{H. \surname{Su}}
\affiliation{%
    Department of Chemical and Environmental Engineering, Yale University, New Haven, CT 06520, USA
}%

\author{W. \surname{Zhang}}
\affiliation{%
    Department of Chemical and Environmental Engineering, Yale University, New Haven, CT 06520, USA
}%

\author{C. \surname{He}}
\affiliation{%
    Department of Chemical and Environmental Engineering, Yale University, New Haven, CT 06520, USA
}%

\author{O. \surname{Romanyuk}}
\affiliation{%
    FZU ‒ Institute of Physics of the Czech Academy of Sciences, Cukrovarnicka 10, Prague 16200, Czech Republic
}%

\author{A. \surname{Paszuk}}
\affiliation{%
    Grundlagen von Energiematerialien, Institut f\"ur Physik, Technische Universit\"at Ilmenau, 98693 Ilmenau, Germany
}%

\author{S. \surname{Hu}}
\affiliation{%
    Department of Chemical and Environmental Engineering, Yale University, New Haven, CT 06520, USA
}%

\author{T. \surname{Hannappel}}
\affiliation{%
    Grundlagen von Energiematerialien, Institut f\"ur Physik, Technische Universit\"at Ilmenau, 98693 Ilmenau, Germany
}%

\title{Photoemission study and band alignment of GaN passivation layers on GaInP heterointerface}

\date{\today}
\begin{abstract}

III-V semiconductor-based photoelectrochemical (PEC) devices show the highest solar-to-electricity or solar-to-fuel conversion efficiencies. GaInP is a relevant top photoabsorber layer or a charge-selective contact in PEC for integrated and direct solar fuel production, due to its tunable lattice constant, electronic band structure, and favorable optical properties.  To enhance the stability of its surface against chemical corrosion which leads to decomposition, we deposit a GaN protection and passivation layer. The n-doped GaInP(100) epitaxial layers were grown by metalorganic chemical vapor deposition on top of GaAs(100) substrate. Subsequently, thin 1–20 nm GaN films were grown on top of the oxidized GaInP surfaces by atomic layer deposition. We studied the band alignment of these multi-junction heterostructures by X-ray and ultraviolet photoelectron spectroscopy. Due to the limited emission depth of photoelectrons, we determined the band alignment by a series of separate measurements in which we either modified the GaInP(100) surface termination or the film thickness of the grown GaN on GaInP(100) buffer layers. On n-GaInP(100) surfaces prepared with the well-known phosphorus-rich (2x2)/c(4x2) reconstruction we found up-ward surface band bending (BB) of 0.34 eV, and Fermi level pinning due to the present surface states. Upon oxidation, the surface states are partially passivated resulting in a reduction of BB to 0.12 eV and a valence band offset (VBO) between GaInP and oxide bands of 2.0 eV. Between the GaInP(100) buffer layer and the GaN passivation layer, we identified a VBO of 1.8 eV.
 The corresponding conduction band offset of -0.2 eV is found to be rather small. Therefore, we evaluate the application of the GaN passivation layer as a promising technological step not only to reduce surface states but also to increase the stability of the surfaces of photoelectrochemical devices.
\end{abstract}
\maketitle

\section{Introduction}
The quest for sustainable energy sources has driven extensive research into semiconductor materials for potential applications in photoelectrochemical (PEC) cells, particularly for direct solar-driven water splitting and other fuel generation \cite{Cheng2018}. Among various semiconductors, III-V compounds play a crucial role in a wide range of electronic and optoelectronic devices, including diode lasers, light-emitting diodes, photodiodes, optical modulators, and multi-junction photovoltaic or PEC devices \cite{Geisz2020,Cheng2018,May2015}.
Ternary compounds, in particular, offer the ability to precisely adjust band gaps and lattice parameters, which allows the flexibility of forming and fine-tuning well-aligned heterojunctions, and thus, to precise control their electronic properties \cite{Ochoa-Martinez2018}. Notably, their use in multi-junction solar cells recently allowed achieving remarkable solar-to-electricity conversion efficiencies of $47.1\%$ \cite{Geisz2020} and, $47.6\%$ \cite{fraunhofer2022develops}.
They are also excellent candidates for solar water splitting, by boosting the exploitation of solar radiation to higher photon-to-electron hydrogen conversion rates than traditional silicon-based systems \cite{green2019solar,Urbain2016}.  The exploration and utilization of III-V compound semiconductors in devices for sustainable energy production have opened up exciting possibilities for more efficient solar energy conversion.

Ga$_{x}$In$_{1-x}$P, is widely acknowledged among III-V semiconductors devices. When grown lattice-matched to GaAs, its band gap is of around 1.8 eV, making it an excellent choice to use as a top photoabsorber in highly efficient tandem structures $20\%$ \cite{May2015, bett2013overview, Feifel2018} with solar-to-hydrogen conversion efficiency surpassing $20\%$ \cite{may2017benchmarking, hu2013analysis, Gu2016}. 
Despite its exceptional performance, Ga$_{x}$In$_{1-x}$P faces significant challenges related to corrosion and its durability when exposed to aqueous solutions and light. Surface corrosion and chemical instability can lead to increased recombination losses, which adversely affect device performance. Consequently, its suitability for immersed operation in a PEC setup is critical \cite{Wang2012, Khaselev1998}. To address this challenge and to enhance the stability of III-V materials during PEC operations, various strategies have been applied, such as utilization of appropriate protection layers as well as chemical and electronic surface and interface passivation \cite{Gu2017, Gu2016, May2015}. By implementing all these elements, researchers aim to improve the performance, stability, and viability of Ga$_{x}$In$_{1-x}$P for its integration into PEC devices.

For example, Gu et al. achieved stable operation of p-GaInP/TiO$_{2}$/cobaloxime for 20 minutes \cite{Gu2016}, while Wang et al. demonstrated stability for up to 70 hours by deposition MoS$_{2}$ on GaInP$_{2}$ \cite{Wang2019}. 

Passivation layers must have larger bandgaps than the one of the top photoabsorber and the charge collector layer to prevent undesirable absorption of the solar light. Moreover, the conduction band offset (CBO) between the protective layer and the photoabsorber should be as small as possible, ideally negligible.  A large CBO, for instance, leads to a reduced conductivity for electrons and hinders their selective transport, which is essential for an optimized device performance \cite{wurfel2014charge, schleuning2022role}. Thus, Understanding the electronic structure and the interface of the photoabsorber and the protective passivation layer is also crucial to driving improvements in performance. Such an interface can reduce non-radiative interfacial recombination as well as promote charge carrier selective transport. 
The determination of valence band offsets (VBOs) through Kraut's method using x-ray photoelectron spectroscopy (XPS) has emerged as a prominent approach in characterizing semiconductor materials \cite{Kowalczyk1983}. Through the analysis of core-level (CL) and valence-band spectra, the energy alignment and band offset between different materials can be accurately assessed.
Kraut's method considers the charge transfer across the interface of two materials. First, the binding energy of the high-intensity core level and the valence band maximum (VBM) is taken into account for both materials. Subsequently, data are obtained from the samples on which the first material is coated with a thin film of the second material. It is crucial that the thickness of the thin film is sufficiently small to allow the photoelectrons from the underlying layer to escape and be detected during the XPS measurement. This allows precise interface-sensitive measurements in which photoemission (PE) peaks from both materials are resolved \cite{Kowalczyk1983, Waldrop1985}.

GaN can serve as an excellent passivation layer for the top III-V photoabsorber used in PEC devices, especially where cathodic reactions, such as hydrogen evolution or carbon dioxide reduction reaction, take place. The band edges of GaN are essential for this specific function \cite{Vanka2018, Tournet2023}. GaN is widely used in electronics due to its ideal properties, e.g. favorable bandgap, for surface protection \cite{Wang2019}. Compared to III-As and III-P compounds, III-nitrides, like GaN, have stronger ionic chemical bonds, which make them more stable over wider pH \cite{Walukiewicz2001, Tournet2023}. Moreover, compared to TiO$_{2}$ ALD-grown protection layer, GaN has a higher stability under a wider range of applied bias \cite{Baker2017, Acevedo2013}. Therefore, GaN is a favorable choice as a functional protecting and passivating layer since it minimally absorbs sunlight, thanks to its wide bandgap of approximately 3.4 eV \cite{Himmerlich2013, AgerIII2004, Walukiewicz2001}. In this regard, the band alignment at the interface of GaN and the photoabsorber, as well as at the GaN surface, plays a critical role in PEC applications. By understanding and optimizing these band alignments, more efficient and stable PEC devices for sustainable energy generation can be designed.

Here, considering GaN's promising properties as a protective layer for III-V photoabsorbers for PEC applications, the band alignment of the GaN/GaInP heterointerface was studied using photoemission spectroscopy (PES). GaN layers with various thicknesses were deposited using atomic layer deposition (ALD), and their band structures were studied with PES, leading to the construction of the final band diagram for the GaN/GaInP heterointerface. This research aims to understand the band alignment at the heterointerface and optimize the properties of the GaN protective layer/III-V photoabsorber interface to enhance the stability and performance of photoelectrochemical cells for sustainable energy generation.


\section{Experimental}

The n-Ga$_{0.51}$In$_{0.49}$P(100) (hereinafter referred to as n-GaInP) epitaxial layers were grown on the n-GaAs(100) substrate, misoriented by 0.1° in the ⟨111⟩ direction, in an H2-based horizontal-flow metalorganic vapor phase epitaxy (MOVPE) reactor (Aixtron, AIX-200). The process was monitored \textit{in situ} with surface-sensitive reflectance anisotropy spectroscopy (RAS, LeyTec) \cite{aspnes1985anisotropies, weightman2005reflection}. The initial surface preparation in MOVPE consisted of deoxidation of the GaAs(100) substrate at 620°C while being exposed to tertiarybutylarsine flow for 10 minutes prior to n-GaInP growth \cite{GaAsdeoxidation}. Next, the growth of a 100 nm thick GaAs(100) buffer layer and a 400 nm thick silicon-doped GaInP buffer layer was carried out at a temperature of 600°C and a pressure of 100 mbar, utilizing tertiarybutylphosphine (TBP), trimethylindium (TMIn), and triethylgallium (TEGa) with a molar V/III ratio of 21.2 and 32.6, respectively. 

The n-GaInP growth rate of 0.24 nm/s was determined on reference samples by \textit{in situ} RAS measurements of the period of Fabry-Perot oscillations \cite{Watatani2005}. The prepared n-GaInP surface was cooled down to 300°C under the flow of TBP precursor. Following this, to prepare the phosphorus terminated (P-rich) $(2 \times 2)$/c$(4 \times 2)$ surface reconstruction, the temperature was ramped up to 310°C, while the TBP was switched off, and RA spectra were measured continuously until the RA signal reached a stable state and constant intensity \cite{Doscher2011}. Ditertiarybutyl silane (DTBSi) was used as the n-type dopant source, with the molar flow adjusted to achieve carrier concentrations of approximately $4.0 \times 10^{18}$ cm$^{-3}$ and $2.0 \times 10^{16}$ cm$^{-3}$ in the n-GaAs homoepitaxial buffer and the n-GaInP layer, respectively.
Electrochemical capacitance-voltage profiling (ECVP, WEP-CVP 21) was used to determine the carrier concentration depth profile in the III-V epilayers (\textit{ex situ} with 0.1 M HCl solution, under visible light illumination at room temperature) \cite{Blood1986}. Results of carrier concentration are shown in the Supplementary Materials (SM) in Fig. S1.

\textit{Ex situ} high-resolution X-ray diffractometry (HR-XRD) scans (Bruker AXS D8 Discover with Ge(022)x4 asymmetric monochromator and Goebel mirror) were used to confirm if the n-GaInP epi-layer is lattice matched to the GaAs(100) substrate. Using XRD observations with diffraction peak maxima at 33.04°, the atomic stoichiometry in the bulk was determined to be $x = 0.51$ in Ga$_x$In$_{1-x}$P composition, which corresponds to the lattice constant of $a = 5.654$ Å, which is the same lattice constant as the GaAs(100) substrate \cite{Tournet2020}. HR-XRD data are included in the SM, Fig. S2.

The samples were transferred from the MOVPE reactor to the ultra-high-vacuum (UHV) analytical chamber by a UHV transfer system, which involved a mobile UHV shuttle operating under a pressure of $10^{-10}$ mbar \cite{Hannappel2004}. The analytical chamber is equipped with XPS (monochromated Al-K$\alpha$ line, Specs Focus 500/Phoibos 150/1D-DLD-43-100), ultraviolet photoelectron spectroscopy (UPS, HeI, Specs Focus 500/Phoibos 150/1D-DLD-43-100), and low-energy electron diffraction (LEED, Specs ErLEED 100-A) \cite{Hannappel2004}.

Prior to the deposition of GaN films the air-exposed surface of n-GaInP was etched using a commercial 1.25 M solution of HCl in 2-propanol (Aldrich) for 1 minute to eliminate weakly bound surface contaminant species. For typical III-V semiconductors, the native oxide layer has been successfully removed using a similar technique \cite{Hajduk2018}. Etching the n-GaInP surface with only a 1 M aqueous HCl solution was also studied, but not applied here, as it was less effective in removing the oxide, leaving a significant quantity of phosphates. Ar-ion sputtering was not used here, since it can change the chemical composition of the surface \cite{Lebedev2016}.

GaN layer deposition was performed using a commercial ALD system (Ultratech Fiji G2). Following the etching of the oxidized n-GaInP, the samples were loaded into the ALD system. Trimethylgallium (TMGa, Strem Chemicals, $99.9999 \%$-Ga) was used as the gallium precursor, and N$_{2}$ plasma served as the nitrogen precursor. Ultra-high purity grade argon and nitrogen gas tanks (Airgas, Inc.) were used as the gas supply for the ALD system. The temperatures of TMGa precursor cylinder, the delivery line, and the susceptor were maintained at 25 $^{\circ}$C, 150  $^{\circ}$C, and 350  $^{\circ}$C respectively. Each ALD cycle consisted of a 0.015 s pulse of TMGa at a flowrate of 30 sccm Ar carrier gas, 80 sccm Ar plasma, and 5 sccm N$_{2}$ with turbo pump disconnected, resulting in a process pressure of 0.12 Torr. This was followed by three 20 s intervals of 300 W plasma at flow rates of 30 sccm Ar carrier gas, 200 sccm Ar plasma, and 40 sccm N2 while the turbo pump was connected, resulting in a  process pressure of 0.03 Torr. The thickness of the GaN ALD coating was determined by the cumulative number of performed ALD cycles. Thin layers of GaN coatings with thicknesses of 1 nm, 3 nm, 5 nm, 10 nm, and 20 nm were deposited by 20, 60, 100, 200, and 400 ALD cycles, respectively. 

XP spectra were measured with respect to the calibrated Fermi level by using sputtered Au, Ag, and Cu reference samples. Spectra were acquired using two photoelectron emission angles: the initial measurements were conducted at a standard emission angle of 90°, and tilting the sample at an angle of 30° to enable a more surface-sensitive analysis. The survey and high-resolution core level spectra were measured with a pass energy of 30 eV and 10 eV, respectively, with an energy step of 0.5 eV and 0.05 eV. UPS He I spectra were measured with an energy step of 0.05 eV. We used the SpecsLab2 and CasaXPS software to analyze and fit each XP spectrum and a Lorentzian function to fit the Ga 2p$_{3/2}$, N 1s, P 2p, and In 3d$_{5/2}$ core level spectra. Using the corresponding fitted PE areas and the relative sensitivity factors, the chemical composition of the film were quantified.

\section{results and discussion}
\subsection{Analysis of the P-rich GaInP(100) surface reconstruction}

\begin{figure*}
  \centering
  \includegraphics[width=1\textwidth]{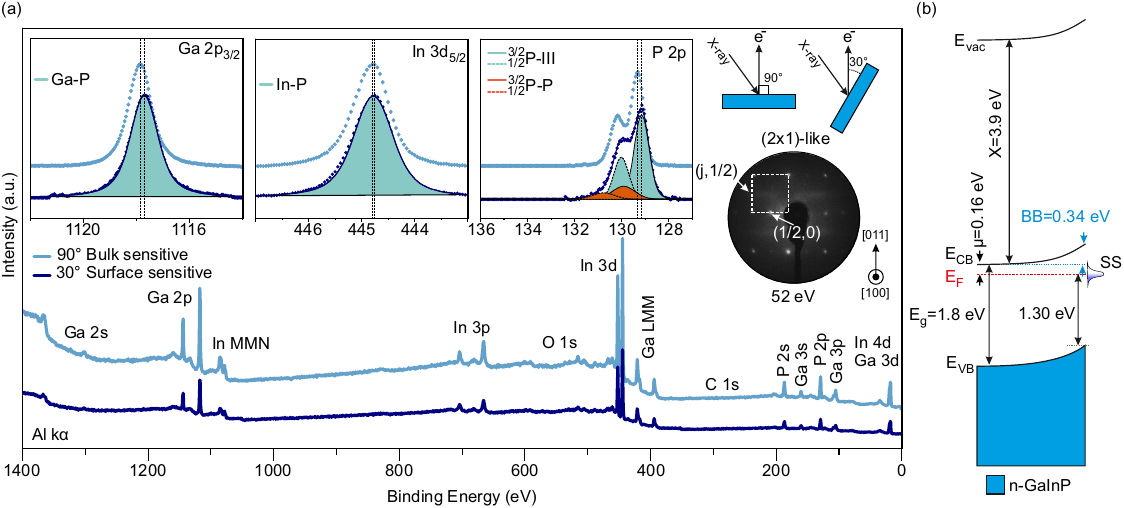}
  \caption{a) The survey and high-resolution XPS spectra of the P-rich n-GaInP(100) surface, accompanied by a LEED pattern demonstrating a  $(2 \times 1)$-like surface reconstruction. Above the LEED pattern, a schematic picture of the experimental setup is shown. At the top: high-resolution spectra of Ga 2p$_{3/2}$, In 3d$_{5/2}$, and P 2p (from left to right) with corresponding fitted data of 30° measurement. Light blue represents measurements at 90°, while dark blue represents measurements at 30°. All spectra are shown after the subtraction of the background. b) ‌Sketch of the band diagram of the P-rich n-GaInP(100) $(2 \times 2)$/c$(4 \times 2)$ surface.}
  \label{fig:P rich survayLEED}
\end{figure*}
Fig.\,\ref{fig:P rich survayLEED}(a) shows, XP spectra and LEED pattern (inset) measured on the as-grown n-GaInP clean surface. The LEED pattern, taken at primary electron beam energies of 52 eV, contains half-order spots and diffused half-order streaks along the [011] direction, thereby indicating a P-rich $(2 \times 1)$-like surface reconstruction \cite{pour2022band}. Analogous LEED patterns were formerly observed on P-rich GaP(100) and InP(100) surfaces \cite{Kleinschmidt2011}, as well as on other ternary compound surfaces such as AlInP \cite{pour2022band}. These surfaces consist of arrays of buckled P-P dimers, each passivated by one hydrogen atom. The saturation of P-P dangling bonds by H induces arrangements of p$(2 \times 2)$ and c$(4 \times 2)$ surface unit cells - the origin of diffuse streaks in the LEED pattern \cite{Kleinschmidt2011}. On an ideal surface, all P-P dimers are passivated by one hydrogen atom. However, it is observed that hydrogen might be missing at P-P dimers which leads to the emergence of defects, which produce surface states in the midgap of the semiconductor that may pin the Fermi level \cite{Moritz2022}. These defects are believed to arise from the desorption of hydrogen atoms from the surface. Moritz et al. have extensively discussed the likelihood of such defects occurring, taking into account variations in temperature and pressure.

The survey spectrum provides clear evidence that there is no surface contamination present. The P 2p core level is fitted with two spin-orbit split pairs shown in Fig.\,\ref{fig:P rich survayLEED}: The dominant pair corresponds to the contribution from the bulk, and the weaker component pair is shifted to a higher binding energy by 0.79 eV (orange). This component is related to the surface bond and provides evidence for the existence of P-P dimers on the surface \cite{Glahn2022}. The surface component becomes larger for a 30° emission angle. In contrast to P 2p, both Ga 2p$_{3/2}$ and In 3d$_{5/2}$ core-level spectra were fitted with only one component related to the bulk Ga-P and In-P bonds, respectively.

Changing the photoelectron emission angle from 90° to 30° induced a shift of all core level peaks by 0.05 eV toward lower binding energy. This is an indication of an upward surface band bending (BB). Considering the band gap of n-GaInP as 1.8 eV \cite{Young2017}, along with an electron affinity of 3.9 eV \cite{Jiang2003}, the measured VBM of 1.30 eV [see Fig.\,\ref{fig: etched Vs Prich-Band D} (d)], the Fermi level position with respect to the conduction band minimum (CBM) of -0.16 eV (see Eq.\,\ref{eq:mu_equation}), and the core level position shifts, the band diagram in Fig.\,\ref{fig:P rich survayLEED}(b) was derived. The magnitude of BB on the P-rich n-GaInP surface was determined to be 0.34 eV. The observed upward band bending in the band diagrams suggests the presence of surface states and Fermi-level pinning \cite{Moritz2022}. Within analogy to a P-rich InP(100) surface, hydrogen-related defects (missing H atom at P-P dimer) introduce states within the bandgap, leading to a pinning of the Fermi level. Given the signals observed in P-rich InP(100), it is anticipated that similar phenomena will be observed on all P-rich III-P surfaces \cite{Moritz2022}.


Conduction band minimum and Fermi level energy difference were estimated using the following equation \cite{wurfel}: 
\begin{equation}
\centering
\mu = -kT \ln \left(\frac{N_C}{n_D}\right)
\label{eq:mu_equation}
\end{equation}
The effective density of states in the conduction band is denoted by \(N_C\), while the doping concentration is represented by \(n_D\).
According to estimates, \(N_C\) for n-GaInP, is 9.76$\times$10\(^{18}\) cm\(^{-3}\), which represents the average of the values for GaP and InP \cite{goldberg1999handbook}. The estimated doping concentration of n-GaInP was approximately 2.0$\times$10\(^{16}\)cm\(^{-3}\) as measured by ECVP [see SM, Fig.\,S1]. 

\subsection{Analysis of the oxidized GaInP(100) surface}

\begin{figure*}
  \centering
  \includegraphics[width=1.0\textwidth]{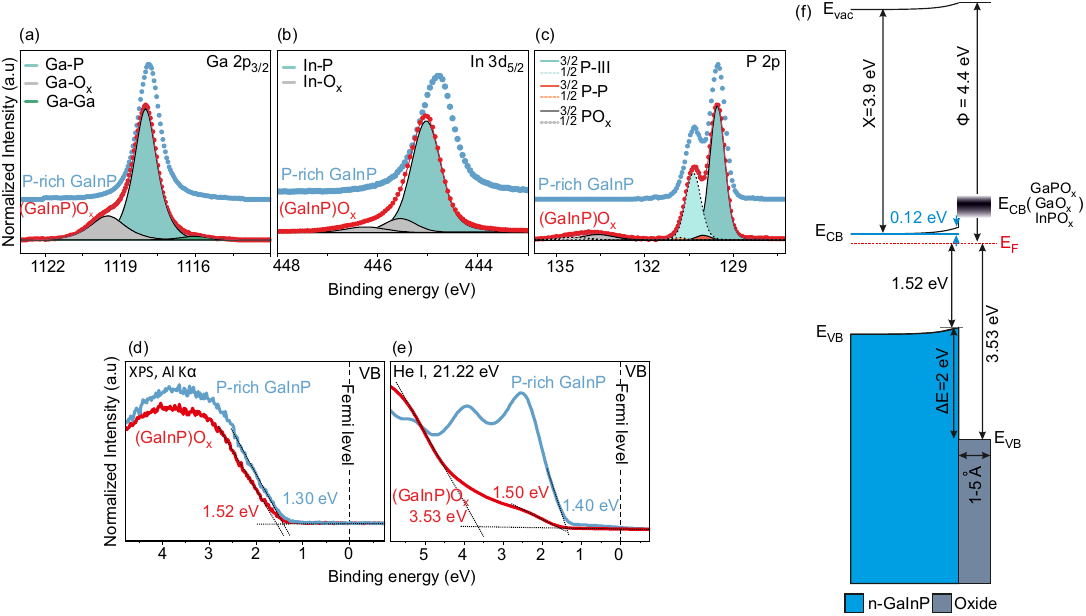}
  \caption{Measured core level spectra of a) Ga 2p$_{3/2}$, b) In 3d$_{5/2}$ and c) P 2p after background subtraction on P-rich n-GaInP (blue) and etched n-GaInP ((GaInP)O$_{x}$, red). (d) The VB-XPS and (e) VB-UPS spectra contain photoelectron contributions from different depths. The measured VBM of the GaInP substrate and topmost oxide layer are indicated.  f) Band diagram of n-GaInP after etching ((GaInP)O$_{x}$).}
  \label{fig: etched Vs Prich-Band D}
\end{figure*}

The air-exposed P-rich n-GaInP surface was wet-chemically etched as described in the experimental to eliminate surface contaminant species, and reduce the oxide layer thickness \cite{Hajduk2018}. Fig.\,\ref{fig: etched Vs Prich-Band D} shows the (a) Ga 2p$_{3/2}$, (b) In 3d$_{5/2}$, and (c) P 2p core level spectra of the as-prepared P-rich n-GaInP surface (blue curves, same spectra shown in Fig.\,\ref{fig:P rich survayLEED}) and the oxidized/etched surfaces ((GaInP)O$_{x}$) (red curves). The shown spectra were measured at 90° angle. Upon comparing the corresponding core level spectra of those two samples, changes in the line shape, as well as core level shifts to higher binding energies, were observed: The spectra of the oxidized samples include contributions from the oxide bonds (gray curves), and the shifts correspond to a reduction of upward surface BB. Thus, the change in the BB indicates a change in charge states on the surface, implying a partial passivation of the surface states.
Apart from the oxide components, on the oxidized sample, in the P 2p spectrum, a small surface component (orange line) at about 129.8 eV is present Fig.\,\ref{fig: etched Vs Prich-Band D}(c). This component could potentially be attributed to a presence of residuals of P-Cl (around $3\%$ Cl) caused by etching in HCl acid or could also be attributed to residuals of the P-P bonds on the surface \cite{hajduk2023photo}.

Fig.\,\ref{fig: etched Vs Prich-Band D}(d) and (e) shows the VB spectra measurements on both, the etched (red curve) and as-prepared P-rich n-GaInP (blue curves) surfaces using XPS and UPS. The information depth of the VBM measured by UPS (VB-UPS) (around 1-2 nm) is smaller than the information depth of VBM measured by XPS (VB-XPS) (5-10 nm). Thus, the utilization of UPS can be employed to investigate the electron emission originating from the native oxide layer, while the emission from the n-GaInP layers superimposes with the emission from the oxide layer in the case of VB-XPS. \cite{hajduk2023photo}. The VBM of the native oxide on the n-GaInP surface is approximately 3.53 eV [Fig.\,\ref{fig: etched Vs Prich-Band D}(e), red curve], while the VBM of n-GaInP visible both by XPS and UPS is 1.52 eV. In the case of the clean P-rich surface, we determined the position of the VBM at about 1.30 ± 0.05 eV [Fig.\,\ref{fig: etched Vs Prich-Band D}(d), blue curve].

Based on the measurements of the VBM, the band diagram of the oxide/n-GaInP surface was constructed in Fig.\,\ref{fig: etched Vs Prich-Band D}(f). The band gap energy of the oxide layer was assumed to be around 4.5-4.8 eV, based on literature values of the corresponding oxide species, including GaPO$_4$, InPO$_4$, and GaO$_x$ \cite{goldberg1999handbook}. Consequently, the lower and upper limits of the CBMs are illustrated by the band gaps of the various species present in the oxide layers. The thickness of the native oxide layer was determined to be ca. 1-5 Å \cite{May2015, plusnin2016atomic} [see SM, Table. S1]. Considering the value of the Fermi level position from the CBM, chemical potential µ of -0.16 eV, and the VBM of 1.52 eV, the band bending of n-GaInP decreased from 0.34 eV to 0.12 eV after oxidation. Note that the absence of band bending within the oxide layer is regarded as a reasonable assumption due to its small thickness and probably the limited presence of mobile charge carriers \cite{hajduk2023photo}. We derived a VBO of 2.0 eV between the oxide and n-GaInP layers. In the following, the oxidized n-GaInP samples were used as substrates for the growth of GaN layers by ALD.

\subsection{Analysis of the GaN/oxide/GaInP(100) heterostructure}

\begin{figure*}
  \centering
  \includegraphics[width=1.0\textwidth]{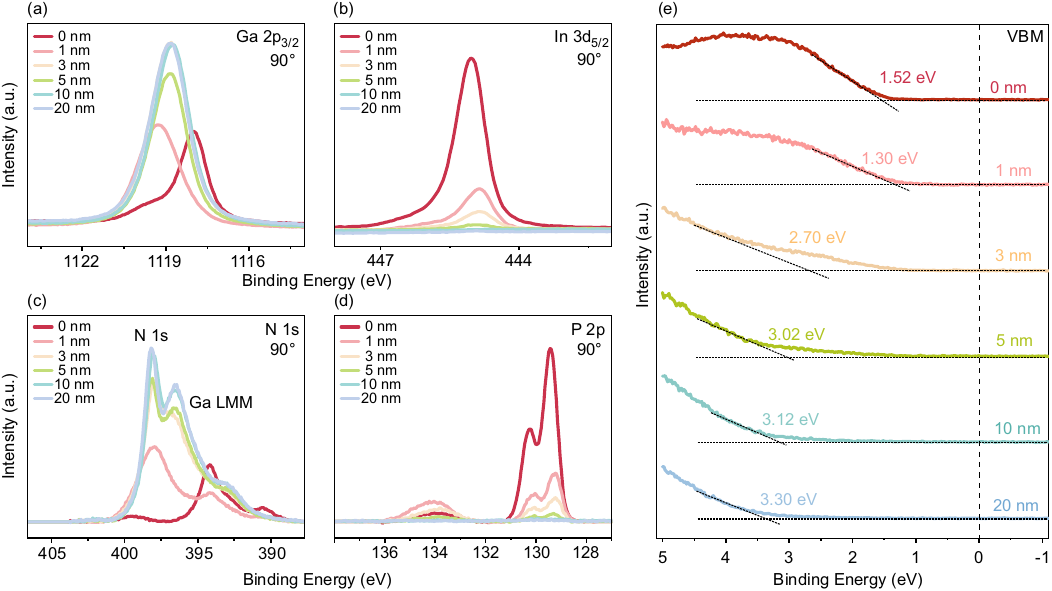}
  \caption{a-d) Core level variations and e) VBM with respect to the thickness of the GaN overlayer.}
  \label{fig: stepwise GaN}
\end{figure*}

A series of GaN layers with a thickness ranging from 1 nm to 20 nm were prepared on n-GaInP samples and were analyzed by XPS and UPS. In Fig.\,\ref{fig: stepwise GaN}(a-d), XPS core level spectra are displayed. With increasing GaN layer thickness, the intensities of the In 3d$_{5/2}$ and P 2p core-level peaks decreased, while the N 1s intensity increased (note that, the N 1s peak intensity for the 0 nm sample is the Ga LMM Auger peak in Fig.\,\ref{fig: stepwise GaN}(c) \cite{Zatsepin2021}). These intensity changes are due to an increase in GaN content on the surface, accompanied by a decrease in photoelectron emission from the n-GaInP substrate.

\begin{figure}[t!]
  \centering
\includegraphics[width=0.5\textwidth]{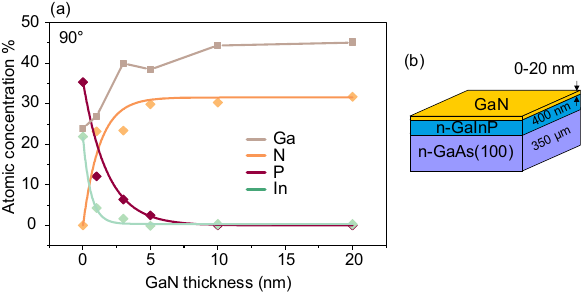}
  \caption{a) The atomic concentration of N, Ga, In, and P of GaN/n-GaInP samples for 0-20 nm thick GaN overlayers. b) Schematic view of the GaN/n-GaInP sample. }
  \label{fig: atomicpercentage}
\end{figure}

Quantification of the samples' atomic composition is shown in Fig.\,\ref{fig: atomicpercentage}(a). The concentrations of Ga and N increased almost exponentially with the rise of GaN layer thickness. It is important to note that the Ga/N ratio is not stoichiometric due to the presence of a Ga oxide layer on the surface. Additionally, fluctuations in the Ga concentration are visible at the beginning of interface formation, which can be attributed to the influence of the thin oxide layer present on the GaInP surface. It should also be mentioned that oxygen and carbon are excluded from this plot but included in the SM, Fig.\,S3. As expected, the concentrations of In and P decreased in parallel to the increase of the GaN thickness. The In concentration dropped more rapidly than the P atomic concentration, which could be related to the higher thickness of the P-based oxides than the In oxides. This could also be an indication of the formation of a formation of a P-containing intermediate layer at the interface rather than an intermediate layer with In.

The Ga 2p$_{3/2}$ peak shows a maximum shift after the deposition of GaN. This shift is attributed to the alignment of the Fermi levels of GaN and n-GaInP, indicating VBO between the two semiconductors with different band gaps. On the samples with a low GaN thickness (1-3 nm), we expect contributions from both the overlayer and the substrate to the Ga 2p$_{3/2}$ peak. Consequently, the PE peak position may change due to varying contributions of the overlayer and substrate at the initial stages of heterointerface formation. However, the Ga 2p$_{3/2}$ and N 1s peak positions remained almost constant on the samples with the GaN thicknesses $>3$ nm. 
A slight binding energy shift of -0.2 eV was observed in In 3d$_{5/2}$ and P 2p core levels on samples with GaN layer thickness between 0 nm and $>1$ nm. This shift suggests a possible charge transfer between the GaN overlayer and the n-GaInP substrate, which could affect the initial oxide/n-GaInP surface band bending. A similar effect was observed during the oxidation of the P-rich surface, as discussed earlier.

\begin{figure*}
  \centering
  \includegraphics[width=1\textwidth]{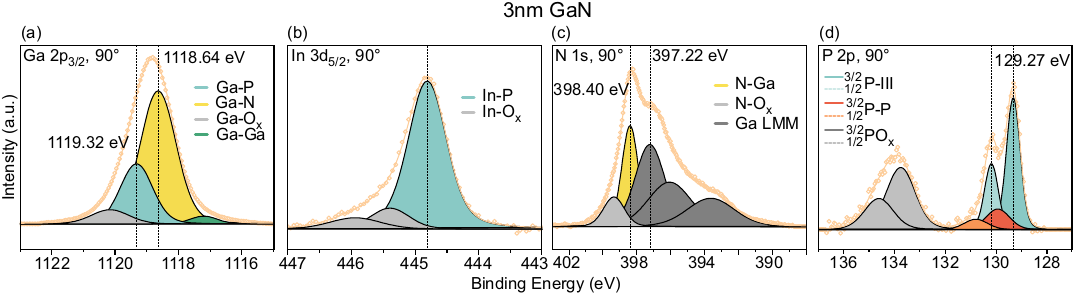}
  \caption{XP spectra of a) Ga 2p$_{3/2}$, b) In 3d$_{5/2}$, c) N 1s, and d) P 2p of 3 nm GaN/n-GaInP sample.}
  \label{fig:3nmGaN}
\end{figure*}

\begin{figure}
  \centering
  \includegraphics[width=0.48\textwidth]{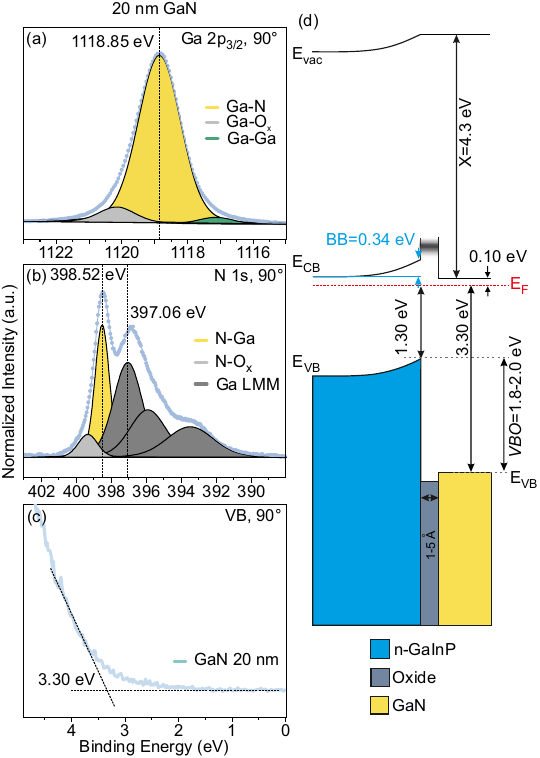}
  \caption{XPS of a) Ga 2p$_{3/2}$ and b) N 1s core level spectra and c) VBM of 20 nm GaN/$n$-GaInP. d) Band diagram of n-GaInP/oxide/GaN. }
  \label{fig:GaN_heterointerface_model}
\end{figure}

Fig.\,\ref{fig:3nmGaN} and Fig.\,\ref{fig:GaN_heterointerface_model} show deconvoluted core level spectra that are measured on thin (3 nm) and thick (20 nm) GaN/oxide/n-GaInP heterostructures, respectively. We can consider the sample with a 20 nm GaN film as only bulk-like since no signal from the GaInP substrate is expected according to the mean inelastic free path of electrons in GaN. In Fig.\,\ref{fig:3nmGaN}(a), the Ga 2p$_{3/2}$ peak of the 3 nm sample contains various contributions to the PE intensity from the GaN overlayer: the Ga-N component at 1118.64 eV, the n-GaInP substrate (Ga-P component) at 1119.32 eV, as well as contributions from the oxide and Ga-Ga bonds. The chemical shift of the oxide component is +1.5 eV, and for the Ga-Ga component is -1.4 eV.

The N 1s spectra of both samples with 3 nm and 20 nm GaN contain contributions from the N-Ga bonds, Ga LMM Auger peaks \cite{Romanyuk2023}, and (Ga)N-O$_{x}$ bonds. It is noteworthy that the binding energy difference between the Ga-N related component positions in the Ga 2p$_{3/2}$ and N 1s peaks is almost the same for both 3 nm and 20 nm: 720.24 eV (1118.64 eV – 398.4 eV) for 3 nm sample and 720.33 eV (1118.85 eV – 398.52 eV) for 20 nm sample. Therefore, the Ga-N component of GaN can be unambiguously identified in the Ga 2p$_{3/2}$ peak in Fig.\,\ref{fig:3nmGaN}(a).

The In 3d$_{5/2}$ peak of the 3 nm sample consists of a component related to In-P and In-O$_x$ bonds, as shown in Fig.\,\ref{fig:3nmGaN}(b). The P 2p peak includes contributions from the P-Ga(In) bonds of the substrate, InP$_{x}$/P-P or P-Cl surface components \cite{hajduk2023photo}, and a pronounced oxide component P-O$_x$. Thus, we confirmed the presence of the intermediate oxide layer at the GaN/oxide/n-GaInP interface and observed no contributions from In or P atoms for the thicker GaN samples. The PE peak positions of all core levels are summarized in Tab.\, I. Comprehensive data fitting for all core levels of the samples with 1-10 nm GaN/n-GaInP can be found in the SM, Fig.\,S4.

The VBO, denoted as $\Delta E_{\text{VB}}$, was calculated using Kraut's equation \cite{Kowalczyk1983, Romanyuk2021}, incorporating the deconvoluted core level peak positions as follows (Kraut's equation):

\begin{equation}\label{Kraut's eq}
\begin{split}
\Delta E_{\text{VB}} = & (E_{\text{P, 2p}}^{\text{GaInP, bulk}} - E_{\text{VBM}}^{\text{GaInP, bulk}}) - (E_{\text{N, 1s}}^{\text{GaN, bulk}} - \\&E_{\text{VBM}}^{\text{GaN, bulk}}) + (E_{\text{N, 1s}}^{\text{GaInP/GaN}} - E_{\text{P, 2p}}^{\text{GaInP/GaN}})  
\end{split}   
\end{equation} 

In Eq. 2, three terms were considered. The first term ($E_{\text{GaInP, bulk P, 2p}} - E_{\text{GaInP, bulk, VBM}}$) represented the energy difference between the PE core level and the VBM of the 0 nm n-GaInP. The second term ($E_{\text{GaN, bulk, N 1s}} - E_{\text{GaN, bulk, VBM}}$) represented the same energy difference but for the 20 nm GaN sample. The third term ($E_{\text{GaInP/GaN, N1s}} - E_{\text{GaInP/GaN, P 2p}}$) included the core level energy difference measured in 1-5 nm GaN/oxide/n-GaInP heterostructures. We used two sets of core level pairs for VBO calculations: the P 2p core level of the substrate (also calculated with In 3d$_{5/2}$, see Tab.\,\ref{table} ) and N 1s core levels of the overlayer. 
The results of these calculations were included in Tab.\,\ref{table}, and the determined VBO values between the n-GaInP substrate and the GaN overlayer were 1.8 ± 0.1 eV.

The resulting band diagram of the n-GaInP/GaN heterostructure is illustrated in Fig.\,\ref{fig:GaN_heterointerface_model}(d). Based on the reported VBO value of 1.8 eV, the VBM at 3.30 eV (see Fig.\,\ref{fig: stepwise GaN}(e)), and the known band gap of GaN (3.4 eV) \cite{Himmerlich2013}, the CBO was determined to be -0.2 eV. This suggests a nearly barrier-free pathway for electron movement across the interface ideal for a selective transport of electrons. In addition, there is a significant VBO, forming a barrier, which reduces the conductivity for holes substantially \cite{wurfel2014charge}.

\begin{table*}
\caption{Measured binding energies and VBOs on $n$-GaInP/GaN by XPS.}
\label{table}
\label{Table 1}
\begin{center}
\begin{tabular}{ |p{0.75cm}||p{1.2cm}|p{1.2cm}|p{1.2cm}|p{1.2cm}|p{1.2cm}|p{1.2cm}|p{0.75cm}|p{1cm}|p{1cm}|}
    \hline
    \multirow{3}{1cm}{\tilt{\textbf{Thicknesses (nm)}}} &
    \multicolumn{6}{c|}{\textbf{Peak position (eV)}} &
    \multirow{3}{1cm}{\textbf{VBM (eV)}}
    &
    \multirow{3}{1cm}{\textbf{VBO (by P 2p$_{3/2}$)}}
    &
    \multirow{3}{1cm}{\textbf{VBO (by In 3d$_{5/2}$)}} \\
    \cline{2-7}
    & \multicolumn{1}{c|}{\textbf{P 2p$_{3/2}$}} &
    \multicolumn{1}{c|}{\textbf{In 3d$_{5/2}$}} &
    \multicolumn{2}{c|}{\textbf{Ga 2p$_{3/2}$}} &
    \multicolumn{1}{c|}{\textbf{N 1s}} & 
    \multirow{2}{1cm}{\textbf{Ga LMM}} & & & \\
    \cline{2-6}
    &
    \multicolumn{1}{c|}{\tilt{\textbf{(P--III)}}} &
    \multicolumn{1}{c|}{\tilt{\textbf{(In--P)}}} &
    \multicolumn{1}{c|}{\tilt{\textbf{(Ga--P)}}} &
    \multicolumn{1}{c|}{\tilt{\textbf{(Ga--N)}}} &
    \multicolumn{1}{c|}{\tilt{\textbf{(N--Ga)}}} &
    &
    &
    &
    \\
    \hline
    \hline
    0 & 129.44 & 445.03 & 1118 & & &394.46 & & & \\
    1 & 129.28 & 444.87 & 1119.43 & 1118.85 & 398.35 & 397.12 & 1.30 & 1.82 & 1.82 \\
    3 & 129.27 & 444.86 & 1119.32 & 1118.64 & 398.40 & 397.22 & 2.70 & 1.88 & 1.88 \\
    5 & 129.31 & 444.88 & 1119.52 & 1118.79 & 398.49 & 397.07 & 3.02 &1.93 & 1.95 \\
    10 & & & & 1118.77 & 398.38 & 397.04 & 3.12 && \\
    20 & & & & 1118.85 & 398.52 & 397.06 & 3.35 && \\
 \hline
\end{tabular}
\end{center}
\end{table*}

The presence of an oxide layer in the heterostructure is confirmed by the observed oxide contribution in the PE core levels, as depicted in Fig.\,\ref{fig:3nmGaN}. The estimated thickness suggests that the oxide layer present in the heterostructure is thin enough to enable electron tunneling, thereby facilitating the movement of electrons across the interface. 
The observed reduction in the VBM of n-GaInP to 1.30 eV indicates an increase in the degree of band bending after the growth of GaN by ALD. Identifying the precise ingredient responsible for this increase is complicated due to the presence of overlapping constituents at the core levels, including Ga-N and Ga-O. This implies that ongoing chemical interactions occur with the oxide layer throughout the ALD process, leading to alterations in its chemical composition.
The comprehensive band energy diagram reveals that the electronic structure of the GaN/oxide/n-GaInP heterointerface could be well-suited as a photocathode for efficient hydrogen reduction reactions \cite{Ulmer2019}. For that, it is crucial to have an electron selective transport through contact with a favorable alignment of the electronic band structure to the adjacent semiconductor. This alignment ensures that the energy levels of the contact in the conduction band closely match those in the adjacent semiconductor. With such alignment, electron currents can flow selectively between the contact and the semiconductor, leading to an optimal device performance \cite{Schleuning2022}.
The energy level interface between GaN and n-GaInP suggests that this specific structure can facilitate electron transport while hindering hole transport. Based on this observation, a hypothesis can be formulated that the combination of n-GaInP and GaN may potentially be a promising candidate for charge-selective contact.

\section{Conclusions}

The investigation of the GaN/oxide/n-GaInP heterostructure using photoelectron spectroscopy has provided valuable insights into the alignment of its electronic band structure across the heterostructure and highlighted the potential benefits of a GaN passivation layer. Through a comprehensive analysis of the MOVPE-prepared P-rich, n-GaInP(100) surface, the oxidized n-GaInP surface, and ALD-deposited GaN layers, we successfully derived the band alignment in the complex heterostructure using photoelectron spectroscopy. Our study revealed a type-II heterostructure with a VBO of about 2.0 eV between the native oxide layer and n-GaInP substrate and a VBO of 1.8 eV (CBO of -0.2 eV) between the oxidized n-GaInP substrate and the GaN passivation layer. Additionally, we observed evidence of changes in charge states and, accordingly, changes in surface band bending at the initial stages of oxide/n-GaInP and GaN/oxide/n-GaInP interface formation.
Based on these findings, we believe that our research could contribute to the development of a protective layer on the cathode side of photoelectrochemical cells and facilitate the formation of a charge carrier-selective contact between the substrate and photoabsorbers.

\section*{Acknowledgements}
We express our gratitude for the financial assistance provided by the National Science Foundation and the German Research Foundation (NSF-DFG project no. HA 3096/19-1). SH and HS gratefully acknowledge the financial support of the National Science Foundation Award No. CBET-2055416. The authors express their gratitude to A. Müller for providing technical assistance throughout the series of experiments.

\bibliography{Mylit}
\end{document}


\maketitle

\section*{Doping level measured by ECVP}

Electrochemical capacitance voltage profiling (ECVP, WEP-CVP 21) was employed to conduct a depth profile analysis of carrier concentration. The etching solution used was 1 M HCl. The surface of the sample yielded a carrier concentration of \(2.0 \times 10^{16}\) cm\(^{-3}\). Notably, the interface between the GaAs buffer layer and GaInP exhibited a carrier concentration of approximately \(2.0 \times 10^{17}\), indicating the occurrence of silicon diffusion from the GaAs layer into the GaInP layer.
\begin{figure}[H]
  \centering
  \includegraphics[width=0.7\textwidth]{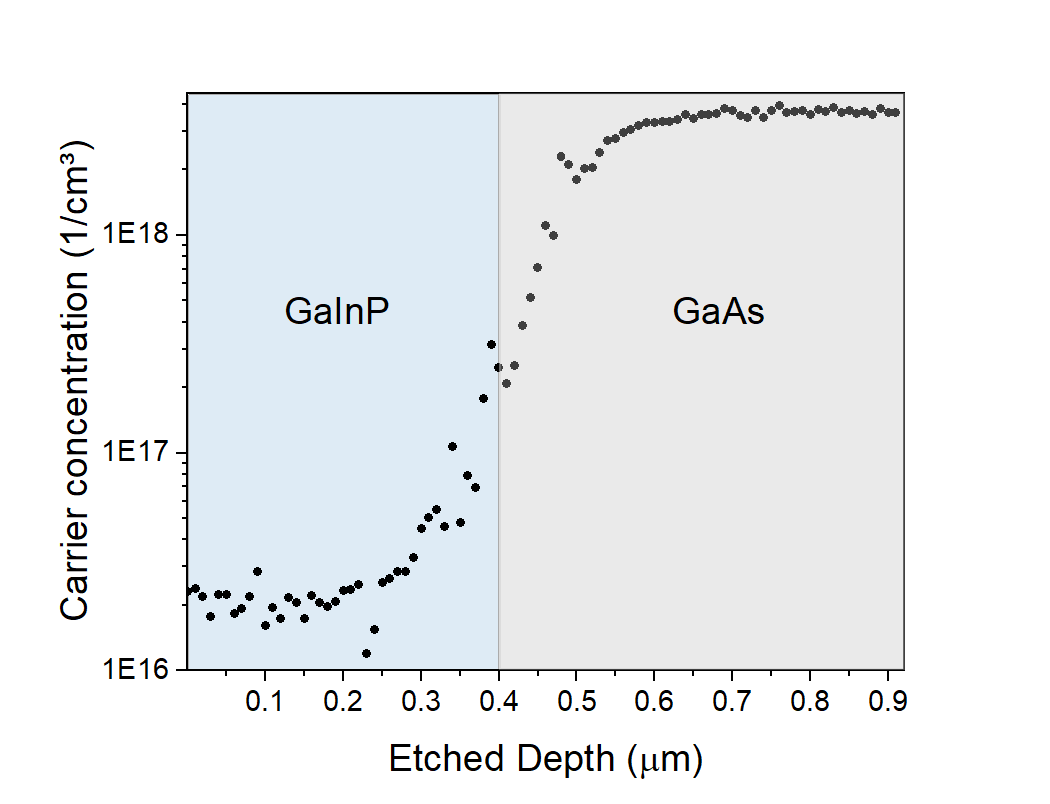}
    \caption*{Fig. S1: Carrier concentration profiles obtained on 400 nm GaInP sample by CVP.}
  \label{fig:dopinglevel}
\end{figure}

\section*{Study lattice matching by XRD measurement}
High-resolution X-ray diffraction (XRD) $\omega/2\theta$ scans of GaAs(004) reflection were measured using a Bruker AXS D8 Discover diffractometer. Fig. S1 displays the diffraction intensity of a 400 nm GaInP layer on GaAs(100). The diffraction peak maxima is located at $33.04^\circ$, confirming an identical lattice constant between the overlayer and the GaAs(100) substrate. Moreover, the derived atomic stoichiometry for indium (In) is $x = 0.51$ in $Ga_{x}In_{1-x}P$.

\begin{figure}[H]
  \centering
  \includegraphics[width=0.7\textwidth]{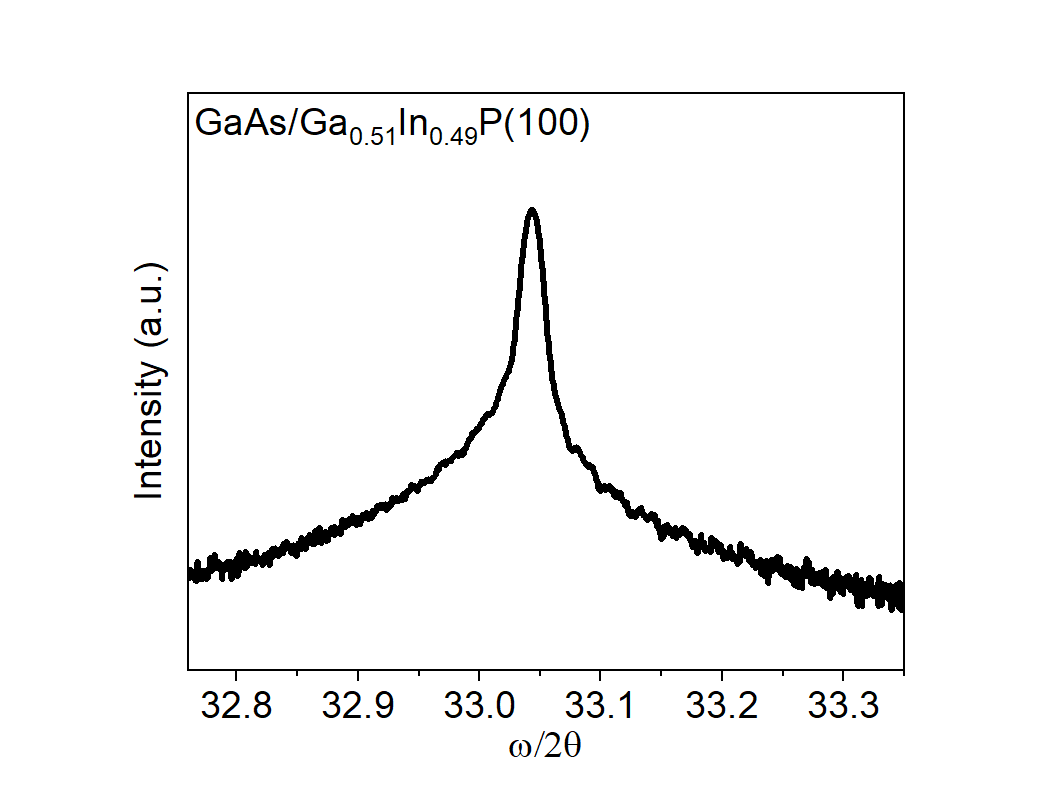}
  \caption*{Fig. S2: XRD scan of GaAs(004) reflection from 400 nm GaInP on GaAs(100) heterostructure.}
  \label{fig:dopinglevel}
\end{figure}

\section*{Thickness of the oxide layer}

To evaluate the oxide thickness, the following equation (Eq. S1) was employed:

\begin{equation}
d = \lambda \ln \left(\frac{I_o}{I_b} + 1\right)
\end{equation}

where \(d\) represents the oxide thickness, \(\lambda\) is the electron attenuation length (taken as 2.3 nm) [1], and \(I_o\) and \(I_b\) denote the intensities of the oxide and bulk, respectively. The resulting oxide layer thickness is estimated to be around 1-5 Å [2].

\begin{table}[h]
\centering
\caption*{Table S1: Calculated oxide layer thickness }
\begin{tabular}{cccccc}
\hline
Thickness & Core levels & Intensity of oxide (I\textsubscript{o}) & Intensity of bulk (I\textsubscript{b}) & Thickness of oxide (Å) \\
\hline
0 nm & P 2p & 1518.61 & 6285.28 & 4.9 \\
       & In 3d5/2 & 1924.58 & 31834.93 & 1.3 \\
       & Ga 2p3/2 & 5765.70 & 31886.91 & 3.8 \\
\hline
\end{tabular}
\end{table}

\section*{Atomic\% of carbon and oxygen}

The atomic percentage of carbon and oxygen with respect to the GaN overlayer thickness is shown in Fig. S3, which was derived from the integrated core level peak intensities with the respective sensitivity factors.

\begin{figure}[H]
  \centering
  \includegraphics[width=0.7\textwidth]{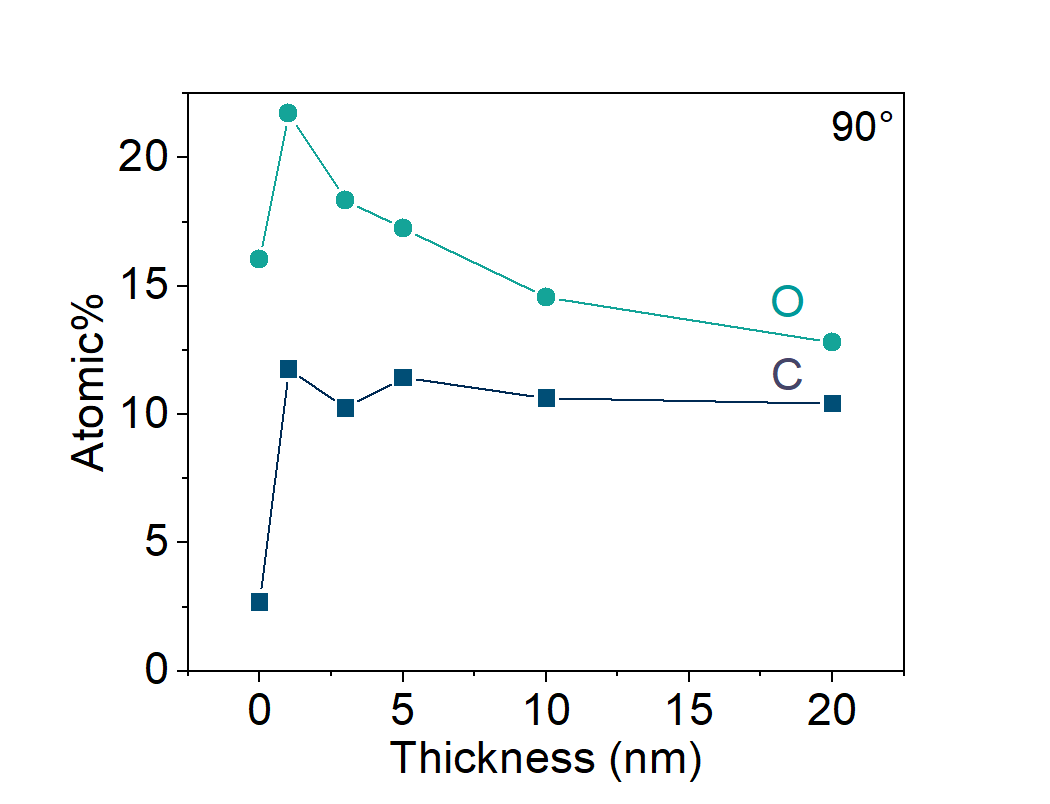}
  \caption*{Fig. S3: The atomic concentration of Oxygen and Carbon of n-GaInP/GaN samples for 0-20 nm overlayer thickness.}
  \label{fig:dopinglevel}
\end{figure}

\section*{Fittings of core levels of 1, 5, and 10 nm GaN/n-GaInP sampels}
The Ga 2p\textsubscript{3/2} peak contains intensity contributions originating from various sources, including the GaN overlayer (Ga-N component), the underlying n-GaInP substrate (Ga-P component), as well as contributions stemming from oxide and Ga-Ga bonds. Within the In 3d\textsubscript{5/2} peak, distinct components are discerned, notably a component associated with In-P bonds and additional features attributable to In-O\textsubscript{x} interactions.
In the N 1s spectra, a range of contributions arise from N-Ga bonds, Ga LMM Auger peaks, and (Ga)N-O\textsubscript{x} bonds.
Notably, a consistent binding energy difference between the positions of the Ga-N related component in the Ga 2p\textsubscript{3/2} and N 1s peaks is observed across all samples, averaging 720.38 ± 0.19 eV. For the P 2p core level, the data is fitted with spin-orbit pairs with the same FWHM and a peak ratio of 2:1, along with a splitting of 0.84 eV. The main component at around 129 eV corresponded to the dominant bulk P-III interaction, and the small feature around 130 eV (orange peak), correlated to the surface InP\textsubscript{x}/P-P or P-Cl components. Additionally, the emergence of a third component around 134 eV indicated the presence of PO\textsubscript{x} species. There were no In or P atom contributions observed for the 10 nm GaN sample.

\begin{figure}[H]
  \centering
  \includegraphics[width=1.0\textwidth]{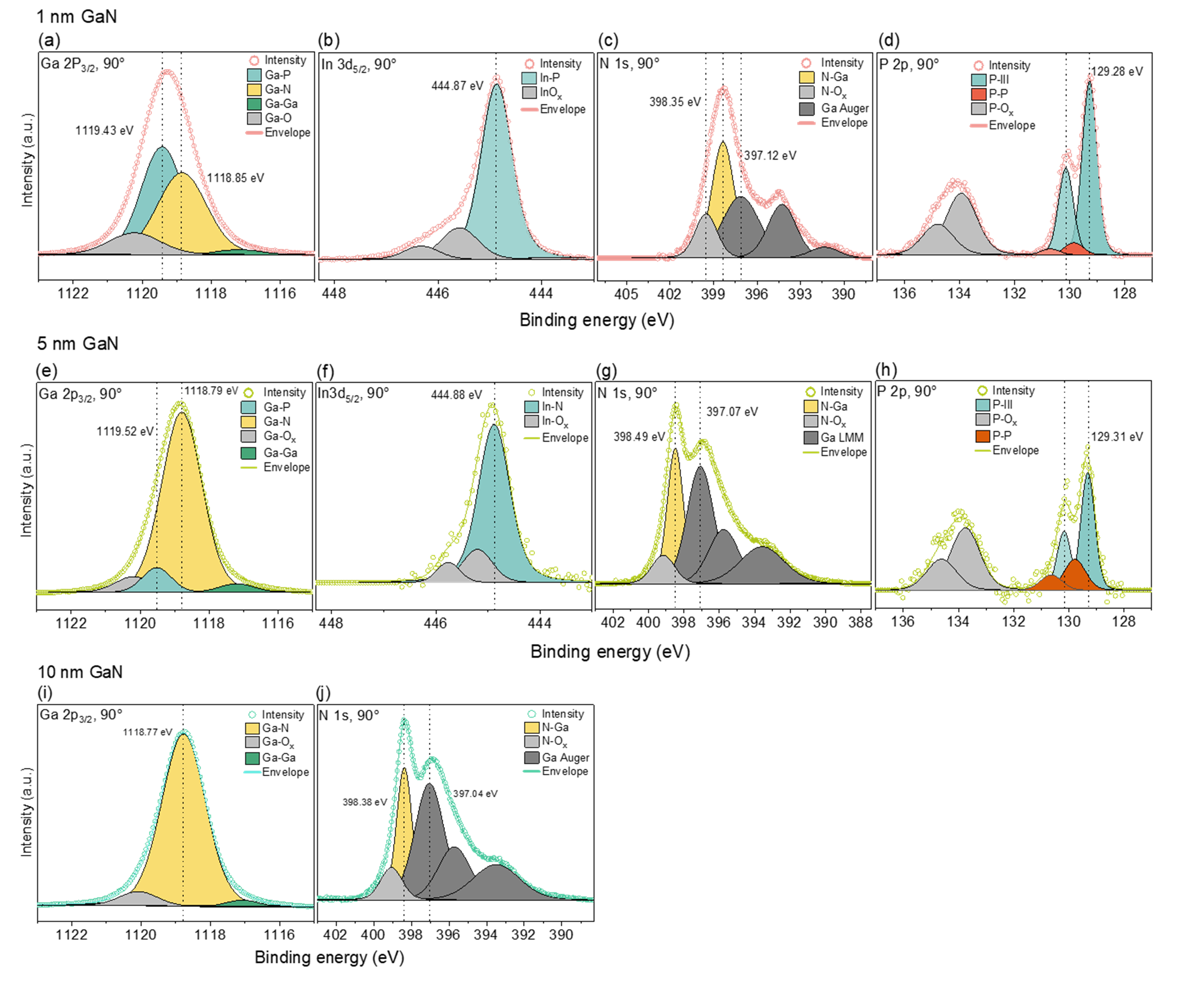}
  \caption*{Fig. S4: XP spectra of a-d) 1nm GaN/n-GaInP, e-h) 5 nm GaN/n-GaInP and i-j) 10 nm GaN/n-GaInP. All spectra are shown after background subtraction.}
  \label{fig:dopinglevel}
\end{figure}

\section*{References}

\begin{enumerate}
    \item Powell, C., Jablonski, A. NIST Electron Effective-Attenuation-Length Database, Version 1.3, Standard Reference Data Program Database 82, National Institute of Standards and Technology, Gaithersburg, MD (2011). URL http://www.nist. gov/srd/nist82.cfm.
    \item Briggs, D., Seah, M. P. Practical Surface Analysis, vol. 1 (Wiley and Sons, Chichester, 1990).
\end{enumerate}

\bibliography{Mylit}